\documentstyle[12pt]{article}

\begin{document}
\renewcommand{\thefootnote}{\fnsymbol{footnote}}
\begin{titlepage}
\null
\begin{flushright}
November 1995 \\
hep-th/9511125
\end{flushright}
\vspace{2.0cm}
\begin{center}
{\Large \bf
BRS COHOMOLOGY IN TOPOLOGICAL STRING THEORY AND
 INTEGRABLE SYSTEMS\footnote{ Talk presented at International symposium
on BRS symmetry, RIMS, Kyoto September 18-22, 1995.}
\par}
\lineskip .75em
\vskip 3em
\normalsize
{\large Hiroaki Kanno}
\vskip 1.5em
{\it Department of Mathematics, Faculty of Science, Hiroshima University, \\
Higashi-Hiroshima 724, Japan}
\end{center}
\vskip3em

\begin{abstract}
In cohomological field theory we can obtain topological
invariants as correlation functions of BRS cohomology classes.
A proper understanding of BRS cohomology which gives non-trivial
results requires the equivariant cohomology theory.
Both topological Yang-Mills theory and topological string theory
are typical examples of this fact.
After reviewing the role of the equivariant cohomology in topological
Yang-Mills theory, we show in purely algebraic framework
how the $U(1)$ equivariant cohomology
in topological string theory gives the gravitational descendants.
The free energy  gives a generating function of topological correlation
functions
and leads us to consider a deformation family of cohomological
field theories. In topological strings such a family is controlled by
the theory of integrable system. This is most easily seen
in the Landau-Ginzburg approach by looking at the contact term interactions
between topological observables.
\end{abstract}
\end{titlepage}

\renewcommand{\thefootnote}{\arabic{footnote}}
\setcounter{footnote}{0}
\baselineskip=0.7cm

\newcommand{\eq}{\begin{equation}}
\newcommand{\en}{\end{equation}}
\newcommand{\eqn}{\begin{eqnarray}}
\newcommand{\enn}{\end{eqnarray}}
\newcommand{\CR}{\nonumber \\}

\newcommand{\del}{\partial}
\newcommand{\cO}{{\cal O}}

\section{Introduction}

The principle of BRS quantization tells us that a BRS exact
term has no contribution to the physical quantities, for
example, the physical $S$ matrix elements. The states
annihilated by the BRS charge are called physical. The observables
should commute with the BRS charge. The expectation values
of observables with respect to the physical states only depend on the
BRS cohomology classes they define.
Hence, the BRS cohomology plays a key role both in conceptual analysis
and in practical computations.
In cohomological quantum field theory the action has topological BRS
symmetry and the notion of BRS cohomology becomes more essential. However,
the observables of our interest (even any BRS closed operators)
are often formally BRS exact
and the naive argument of decoupling of BRS exact operators
prevents us from obtaining topological invariants we expect.
This fact is closely related to the nature of cohomological field theory
sometimes claimed as no propagating degrees of freedom in the theory.
One of ways, in our opinion, to manage this subtle business is to employ
the equivariant cohomology theory \cite{AB,MQ,Kal,Sto}.
(See also an excellent lecture note \cite{CMR}.) The main aim of this article
is to show how the equivariant cohomology in topological
string gives a rich spectrum of the gravitational descendants
and how the contact term interactions between them naturally
give rise to the integrable system which governs the theory.

Let us begin with reviewing some basic structures of cohomological
quantum field theories \cite{W1}.
The property which characterizes cohomological
field theories is that the energy momentum tensor is BRS exact;
\eq
T_{\mu\nu}=\big\{ Q_B, \Lambda_{\mu\nu} \big\}~,
\en
where $Q_B$ is the topological BRS charge.
The observables $\cO_I$ are defined to be $Q_B$ cohomology classes as usual.
If the vacuum is annihilated by the BRS charge, we can see
by the standard argument in BRS quantization procedure the vacuum
expectation value of topological observables
$\langle \cO_{I_1}\cO_{I_2}\cdots \cO_{I_n}\rangle$ defines a (possible)
topological invariant in the sense that it is independent of the
background metric.
Here we have used the fact that the variation of correlation functions
with respect to the background metric is obtained by the insertion of
the energy momentum tensor. It is crucial in this argument that
the topological BRS symmetry is not spontaneously broken.
We require that
\eq
\langle 0| [Q_B, \chi ] | 0 \rangle = 0~. \label{SSB}
\en
But we have to be careful about what kind of the operator \lq $\chi$ \rq\ in
(\ref{SSB}) is ``admissible'', when we use it.
This is the point which is answered by the equivariant
cohomology.

Introducing a parameter $t_I$ for each observable $\cO_I$,
we can define a generating function of topological correlation functions;
\eqn
F[t_I]~&=&~\Bigl\langle \exp (\sum_I t_I \cO_I) \Bigr\rangle~,\CR
\langle \cO_{I_1}\cO_{I_2}\cdots \cO_{I_n}\rangle&=&
\frac{\del^n F}{\del t_{I_1}\del t_{I_2}\cdots \del t_{I_n}}~.
\enn
Explicit examples are discussed below.
The generating function $F[t_I]$ organizes the topological invariants
in a nice way. For example, by deriving recursion relations among
topological correlation functions
one may get a deep insight which is hard to see
for those who look at each invariant separately.
Such topological recursion relations or topological Ward identities
are expressed in terms of differential equations for $F[t_I]$
which sometimes allow geometrical interpretation.
Furthermore, it is possible to think of the generating function as
the path integral by the action with local deformations of the form
${\cal L} \longrightarrow {\cal L} + \sum t_I \cO_I$ .
This means that topological invariants are generated by
the deformations of {\it local} Lagrangian. The fact that
the topological invariants have local density is
quite important, since we will make full use of the machinery of
{\it local} quantum field theory in the path integral
computation of topological invariants.
The most intriguing aspect of cohomological quantum field  theory
is that we have a description of global topological quantities
in the framework of local quantum field theories. In this way,
we are naturally led to the geometry of deformation family of cohomological
field theories. One of the reasons the BRS approach is so natural
and powerful in cohomological field theory is that we are
looking at the deformation theory of {\it local} quantum field theories.
Historically, the BRS approach enjoyed its first success in the renormalization
of non-abelian gauge theory. The viewpoint that
the theory of renormalization is a kind of the deformation theory suggests that
one of the common aspects in BRS approach to local quantum field theory
is the idea of deformation,
where the mathematical concept of cohomology is useful.
A nice and typical example in mathematics is the Kodaira-Spencer theory
of the deformation of complex structure.

\section{Topological Yang-Mills theory}

To illustrate the idea of the equivariant cohomology, we first
look at 4-dimensional topological Yang-Mills theory briefly.
The topological BRS transformation is given as follows \cite{BS,LP};
\eqn
\delta A_\mu &=& \psi_\mu - D_\mu c~, \quad \delta\psi_\mu =
-D_\mu \phi + [\psi, c]~, \CR
\delta c &=& \phi - \frac{1}{2} [c,c]~, \quad \delta \phi = [\phi, c]~,
\enn
where $A_\mu$ is a gauge field and $c$ is the Faddeev-Popov ghost.
The remaining ghost fields $\psi_\mu$ and $\phi$, which are absent
in the physical Yang-Mills theory, are topological BRS
partners of $A_\mu$ and $c$, respectively, as the transformation law indicates.
This transformation law can be identified as the canonical coboundary operator
on the Weil algebra of the universal moduli space \cite{Kan}. From the second
Chern class of this Weil algebra
we can construct a series of operators;
\eqn
\cO^{(0)}&=&\frac{1}{8\pi^2} \hbox {Tr} \phi^2~, \quad
\cO^{(1)}=\frac{1}{4\pi^2} \hbox {Tr} \phi\psi~, \CR
\cO^{(2)}&=&\frac{1}{4\pi^2} \hbox {Tr} (\phi F_A
+\frac{1}{2}\psi\wedge\psi)~, \\
\cO^{(3)}&=&\frac{1}{4\pi^2} \hbox {Tr} \psi \wedge F_A~, \quad
\cO^{(4)}=\frac{1}{8\pi^2} \hbox {Tr} F_A\wedge F_A~,  \nonumber
\enn
which satisfies the descent equation;
\eqn
d \cO^{(4)}&=& 0 \CR
\delta \cO^{(n)} + d \cO^{(n-1)}&=& 0 \quad (1\leq n \leq 4) \\
\delta \cO^{(0)} &=& 0~. \nonumber
\enn
The operator $\cO^{(n)}$ is a space-time $n$-form with ghost number $(4-n)$.
For a simply connected four manifold
topological observables are $\cO^{(0)}(x)$ and
\eq
I(\Sigma) = \int_\Sigma \cO^{(2)}~,
\en
where $\Sigma \in H_2 (M, {\bf Z})$ is a closed two dimensional
surface. The descent equation implies that the topological
correlation function $\langle I(\Sigma_1)I(\Sigma_2)
\cdots I(\Sigma_n)\rangle$ only depends on the
homology class of $\Sigma_i$. It is these correlation functions
which give a field theoretical realization of the cerebrated
Donaldson polynomials.

Now our problem is that $I(\Sigma)$ looks
formally BRS exact, due to the identity
\eq
8\pi^2 \cO^{(2)}= \delta \bigl[ \hbox {Tr} (A\wedge \psi + c dA) \bigr]
+ d \bigl[ \hbox {Tr} (c\psi + \phi A - \frac{1}{2}[\phi,\phi]A) \bigr],
\label{CS}
\en
which follows from the triviality of cohomology of the Weil
algebra. Hence, if we believe in the naive decoupling of BRS
exact operators, we lose the Donaldson polynomials.
However, the idea of the equivariant cohomology saves the situation.
In the equivariant theory we restrict the BRS operator
on the space of the ``basic'' cochains \cite{OSB}, which satisfy;
\eq
{\cal L}_V \chi = \iota_V \chi = 0~. \label{BAS}
\en
More precisely, in the $G$-equivariant theory we have two actions
of the Lie group $G$, given by the Lie derivative
${\cal L}_V$ and the interior product $\iota_V$. The cochain space
of the equivariant cohomology is defined to be the fixed points of both
actions. Together with the exterior derivative (the BRS operator
in our context), ${\cal L}_V$ and $\iota_V$ constitute
basic operations in Cartan's differential calculus.
In the next section we will see that these operations are naturally
accommodated in (topological) string theory.
The first term of the right hand side of eq.(\ref{CS}) is not an admissible
operator in the equivariant cohomology, because
\eq
\iota_V \hbox {Tr} (A\wedge \psi + c dA) = \hbox {Tr} (VdA)~, \label{FP}
\en
where $V$ is a vector field along the orbit of the gauge transformation
group $\cal G$. We should take the $\cal G$-equivariant cohomology
of BRS operator in topological Yang-Mills theory.
In eq. (\ref{FP}) we have followed the usual geometrical identification of
the Faddeev-Popov ghost and regarded it as a basis of one forms
(the Maurer-Cartan forms) along the gauge orbit.

The topological observable $I(\Sigma)$ is
non-trivial in the $\cal G$-equivariant cohomology.
The generating function $F[t_I]$ of topological
invariants in topological Yang-Mills theory is
\eq
F[\alpha_a, \lambda]~=~\Bigl\langle \exp \big( \lambda \cO^{(0)}(x)
+\sum_{a=1}^{b_2}~\alpha_a I(\Sigma_a) \big) \Bigr\rangle~,
\en
where $b_2$ is the second Betti number and $\{\Sigma_a\}$ is
a basis of the second homology group of the four manifold.
One of the most striking
recent developments in topological Yang-Mills theory
is that $F[\alpha_a, \lambda]$ is evaluated
exactly for a large class of four manifolds (called simple type) \cite{W2}.
The expression of $F[\alpha_a, \lambda]$ involves
the classical topological data encoded in the intersection form
on $H_2(M, {\bf Z})$ and the number of the solutions to
the (abelian) monopole equation as a new \lq\lq quantum\rq\rq\ information.
To answer the questions from the side of the traditional
quantum field theory, for example, the issue of spontaneous
symmetry breaking of topological symmetry \cite{Bau}, it is desirable
to have a concept of states associated with four manifolds
(with boundary). In string theory we know such a construction
of topological states for (punctured) Riemann surfaces
in terms of the Universal Grassmannian. In the computation of
$F[\alpha_a, \lambda]$ mentioned above
there appear distinguished two dimensional
homology classes called the Seiberg-Witten class. These classes
regarded as two dimensional world sheets are \lq\lq
cosmic strings\rq\rq\  embedded in the four manifold \cite{W3}.
Hence, four manifolds with string world sheets playing the role
of generalized punctures may be natural
objects in trying to construct topological states in four dimensions.
In any case it would be interesting
to see if the recent progress in understanding four
dimensional topological theory and its cousins gives
some clues in this direction.

\section{Topological string theory}

Topological string theory has the following symmetry of topological
conformal algebra, which can be obtained by twisting the $N=2$
superconformal algebra \cite{EY,DVV};
\footnote{The bracket is ${\bf Z}_2$ graded. It means
anti-commutator, if both entries are fermionic.}
\eqn
\big[L_m, L_n\big]&=&(m-n)L_{m+n}~, \quad
\big[L_m, G_n\big] = (m-n) G_{m+n}~, \CR
\big[L_m, Q_n\big]&=& -n Q_{m+n}~, \quad
\big[L_m, J_n\big] = -n J_{m+n} -\frac{d}{2} m(m+1) \delta_{m+n,0}~,\CR
\big[J_m, Q_n\big]&=&Q_{m+n}~,  \quad
\big[J_m, G_n\big] = -G_{m+n}~, \quad
\big[J_m, J_n\big] = dm \delta_{m+n,0}~, \\
\big[Q_m, G_n\big]&=&L_{m+n} + mJ_{m+n} + \frac{d}{2}m(m+1)\delta_{m+n,0}~.
\nonumber
\enn
We have the same algebra in the anti-holomorphic sector, which
will be denoted with bar in the following. The holomorphic sector and
the anti-holomorphic sector (anti-)commute each other. The constant
$d$ is a $U(1)$ current anomaly and it is the central extension of the algebra.
We identify $Q(z)= \sum Q_{-n} z^{n-1}$ as topological BRS current
and $J(z)= \sum J_{-n} z^{n-1}$ as ghost number current.
The BRS charge is defined by $Q_B = \oint dz Q(z) +\oint d\bar z
\overline{Q}(\bar z)$.
Now we observe the crucial identity
\eq
T(z)= \big[ Q_B , G(z) \big]~, \quad Q(z) =  \big[ J(z), Q_B
\big]~,
\en
which justifies the name ``topological''.  The energy momentum tensor
$T(z)= \sum L_{-n} z^{n-2}$ is topological BRS \lq\lq daughter\rq\rq\ of
the super current $G(z)= \sum G_{-n}z^{n-2}$. It is curious to compare
the first relation with the basic relation in the (bosonic) string theory;
\eq
T^{tot}(z) = [Q_{Vir}, b(z)]~,
\en
for the Virasoro BRS charge $Q_{Vir}$ and the reparametrization
anti-ghost $b(z)$, which is not to be confused with the Nakanishi-Lautrup
field. The total energy momentum tensor $T^{tot}$
has a ghost contribution, so that the total
Virasoro central charge vanishes. This common algebraic
structure leads us to several observations which suggest that
any string theory is topological in some sense.
Furthermore, as several authors have noticed before this algebraic
structure is a stringy version of the Cartan's differential
calculus. We recognize the correspondence;
\eq
(d, {\cal L}_V, \iota_V) \longleftrightarrow (Q_{Vir}, T^{tot}(z), b(z))
\longleftrightarrow (Q_B, T(z), G(z))~. \label{COR}
\en

 A model of topological string theory is obtained
by taking a topological matter theory (a twisted $N=2$ superconformal
model) coupled to topological gravity.
The essential part of topological gravity is the topological Virasoro
ghost system $(c,b,\gamma,\beta)$ which determines
the \lq\lq measure\rq\rq on the
moduli space. The commuting ghosts $(\gamma, \beta)$ are topological
BRS partners of the Virasoro ghosts $(c,b)$.
Concerning the two dimensional metric variables,
there are several options of a field theoretical
realization \cite{LPW,MS,MP,W4,VV,Bec}.
The spectrum of the theory
consists of the primaries, which come from the chiral ring of
the $N=2$ theory, and their gravitational descendants.
The dressing operator which creates the descendants from the primaries
arises from topological gravity sector. It is a
field theoretical realization of the non-trivial cohomology
class on the moduli space of (punctured) Riemann surface
and is an analogue of $I(\Sigma)$ in topological Yang-Mills
theory which represents the non-trivial cohomology class
on the instanton moduli space.
After coupling to gravity the topological BRS charge should be
\eqn
\widehat Q_B &=&Q_{Susy} + Q_{Vir}~, \CR
 Q_{Vir} &=&~\oint c(z)\widetilde{T(z)} -\oint \gamma(z)\widetilde{G(z)}~.
\enn
The gauge part $Q_{Vir}$ with $\widetilde{T(z)}=T_{matter}+ T_{metric} +
\frac{1}{2}T_{ghost}$ and $\widetilde {G(z)} =
G_{matter}+ G_{metric} +\frac{1}{2}G_{ghost}$
is the canonical BRS operator for the (super) Virasoro
algebra. The super charge $Q_{Susy}$ comes from
the (twisted) $N=2$ super symmetry
which is present in any two dimensional topological theory.
Here and below only the holomorphic sector is explicitly written.
But the same expressions with bar are applied
to the anti-holomorphic sector too.

Using the trick of similarity
transformation \cite{Kat}, or constructing an appropriate homotopy
operation, we can find a representative of descendants
which only contains the matter degrees of freedom \cite{EKYY}.
This has an advantage that we do not have to rely on explicit realizations
of topological gravity and enables us to use twisted $N=2$
models to investigate topological string.
The homotopy operation we employ is
\eq
U~=~\exp \biggl(  - \oint dz c(z) \widetilde {G(z)}\biggr)~.
\en
This homotopy transformation simplify the BRS transformation;
\eq
U~(Q_{Susy} + Q_{Vir})~U^{-1}~=~Q_{Susy}~.
\en
at the expense of the following shift in the Virasoro anti-ghost $b(z)$;
\eq
U b U^{-1}~=~b + G^{tot}~.
\en
The appearence of the super current $G(z)$ plays a crucial role in the
following
argument of the equivariant cohomology in string theory.
After coupling to topological gravity, we have to look at
the equivariance condition. Originally this is imposed on
the topological ghost sector and called the semi-relative
condition in the closed string theory;
\eq
(b_0)^- \vert \hbox{state}\rangle = 0~,
\en
where the superscript $^-$ means taking the difference of
the holomorphic sector and the anti-holomorphic sector, {\it e.g.}
$(b_0)^- = b_0 - \bar b_0$.
The semi-relative condition is responsible for the non-decoupling
of the dilaton vertex operators in the closed string theory \cite{DN}.
The symmetry in question is a change of the base point of
the parametrization of the closed string by $S^1$.
It is a $U(1)$ symmetry which is present in any closed string
theory. Now if we perform the similarity transformations
introduced above, the semi-relative condition is transformed into
\eq
(b_0 + G_0)^- \vert \hbox{state}\rangle = 0~.
\en
For the states depending only on the matter degrees of freedom,
we get
\eq
(G_0)^- \vert \hbox{state}\rangle_{matter} = 0~. \label{EQU}
\en
Taking the correspondence (\ref{COR}) into account, we see this
is nothing but the condition of \lq\lq basic\rq\rq\ cochains
(cf. eq.(\ref{BAS})) for the $U(1)$ equivariant cohomology in string theory.

A convenient realization of topological matter is provided by
the Landau-Ginzburg description, which is a powerful tool
in a classification of $N=2$ superconformal models as fixed
points of renormalization group flow and also in a construction of
superstring vacua \cite{Mar,VW,GVW}.
We will use the Landau-Ginzburg model to examine the above idea.
The integrable structure is most easily seen in the Landau-Ginzburg
approach as we will show below.
In superspace the action takes the following form;
\eq
{\cal L}~=~\int d^2 z d^4 \theta~K(X_A, \overline{X_A}) +
\int d^2 z d^2 \theta_+~W(X_A) +
\int d^2 z d^2 \theta_-~\overline{W}(\overline{X_A}) ~,
\en
where the chiral superfields $X_A(z,\theta_+)$
are treated as the Landau-Ginzburg variables.
Due to the non-renormalization theorem, the model is characterized
by the superpotential $W(X_A)$. Let us introduce the following
notations for the component fields;
\eqn
X_A &=& x_A + \theta_+ \psi_A + \bar\theta_+\overline{\psi_A}
+ \theta_+\bar\theta_+  F_A~, \CR
\overline{X_A} &=& \overline{x_A} + \theta_- \rho_A +
\bar\theta_-\overline{\rho_A}
+\theta_-\bar\theta_- \overline{F_A}~.
\enn
Eliminating the auxiliary fields by the equation of motion;
\eq
F_A = \bar\del_A \overline{W}:= \frac{\del \overline{W}}{\del
\overline{X_A}}~,\quad
\overline{F_A} = \del_A W := \frac{\del W}{\del  X_A}~,
\en
we arrive at the action in component fields;
\eq
{\cal L} = \int d^2 z \biggl(
\vert \del x_A \vert^2 + \psi_A \bar\del \rho_A + \overline{\psi_A}
 \del\overline{\rho_A}
+ \vert \del_A W \vert^2 + (\del_A\del_B W) \psi_A\overline{\psi_B}
 + (\bar\del_A
\bar\del_B \overline{W}) \rho_A \overline{\rho_B}
\biggr)~.
\en
After topological twisting which changes the spin of fermions, $(\psi_A,
\overline{\psi_A})$ are zero forms and $(\rho_A, \overline{\rho_A})$
are one forms on the world sheet. The topological BRS transformations are
\cite{Vaf};\footnote{There is another type of BRS transformations in
topological
Landau-Ginzburg model which looks better in some respects. I am grateful
to F.~De Jonghe to point it out in the symposium.}
\eqn
\delta~x_A &=& 0~, \quad \delta~\overline{x_A}
= \psi_A + \overline{\psi_A}~, \CR
\delta~\psi_A &=& \del_A W~, \quad \delta~\overline{\psi_A} =
- \del_A W~, \\
\delta~\rho_A &=& - \del x_A~, \quad \delta~\overline{\rho_A}
= - \bar\del x_A~. \nonumber
\enn
Though the action is {\it not} BRS exact, the energy-momentum tensor
is BRS exact;
\eq
T_{zz}~=~-\delta \Big( \del\overline{x_A}\rho_A \Big)~.
\en
Hence the super-current of the Landau-Ginzburg model is identified with
\eq
G_{zz}~=~-\del\overline{x_A}\rho_A ~.
\en
We note that the super-current $G_{zz}$ is independent of the
super potential. On the other hand the BRS current does
depend on the potential.

Now let us consider the most simple example of $A_k$ type super
potential;
\eq
W = \frac{1}{k+2}~X^{k+2}~, \quad (\del W = X^{k+1})~,
\en
which has a single Landau-Ginzburg variable $X$ and
describes the deformations of topological minimal
models. (For multi-variable case, our understanding of
integrable structure is still poor.) The primary
fields, which coincide with the chiral ring of
the $N=2$ theory, are
\eq
{\cal R}~=~\Big\{ 1, X, X^2, \cdots, X^k \Big\}~. \label{ring}
\en
After coupling to topological gravity we have to
look at the equivariance condition (\ref{EQU}).
Since the super current $G_{zz}$ has conformal weight two, its
zero mode is
\eq
G_0 = \oint dz z \Big( \del\overline{x}\rho \Big)~.
\en
For any polynomial $P(x)$ in $x$, we see
\eq
\Bigl[ G_0, P(x) \Bigr] = 0~,
\en
since we have only simple pole in operator product expansion.
If the polynomial has the form $P(x)= \del W \cdot Q(x)$,
then it is BRS exact;
\eq
P(x) = \Bigl[ Q_B, (\psi-\overline{\psi})Q(x) \Bigr]~.
\en
In fact this is the reason we get the chiral ring (\ref{ring}) before
coupling to gravity.
Now we have
\eq
\Bigl[ G_0, (\psi-\overline{\psi})Q(x) \Bigr] \neq 0~, \label{DP}
\en
since the double pole is created in the operator product expansion in this
case.
The additional contribution comes from $\phi-\rho$ contraction.
Thus, the higher order polynomials are not BRS exact in the $U(1)$ equivariant
cohomology in topological string.
We can check they actually define the same cohomology class
as the gravitational descendants. In the picture we have introduced
above the coupling to topological gravity is effectively achieved by
imposing the equivariance condition on topological matter sector.
Comparing eq.(\ref{FP}) in topological Yang-Mills theory with
eq.(\ref{DP}) above, we see that the observable $I(\Sigma)$ for the
Donaldson polynomials and the gravitational descendants,
typically the dilaton operator in string theory, have the common
feature.

\section{Integrable Deformation}

We have seen that the $U(1)$ equivariant cohomology in topological
string theory gives us a rich spectrum of the gravitational descendants.
For each primary field $\phi_\alpha,~(\alpha = 1,2, \cdots, \hbox {dim.}
{\cal R})$, we have a tower of its descendants
denoted by $\sigma_n (\phi_\alpha),~(n = 1, 2, \cdots)$.
In the Landau-Ginzburg description of the minimal models
these descendants correspond to higher order polynomials in the Landau-Ginzburg
variable. We will show more precise identification below.
Following a general prescription, we introduce the free energy of
topological string;
\eq
F[ t_{\alpha, n} ]~=~\Bigl\langle \exp \Big( \sum_{\alpha, n} t_{\alpha, n}
\sigma_n (\phi_\alpha) \Big) \Bigr\rangle~,
\en
where $n=0$ part stands for the primaries.
A standard way of associating an integrable structure with
topological string is to claim that $F[ t_{\alpha, n} ]$ is the logarithm
of a tau function of KP or Toda lattice hierarchy \cite{Dij}.
The basic generator of the $U(1)$ equivariant cohomology allows us
to introduce a spectral parameter and the gravitational descendants
are identified with the Hamiltonians of higher integrable flows.
In the Landau-Ginzburg approach the super potential
is naturally promoted to (the dispersionless limit of) the Lax
operator. This observation makes the Landau-Ginzburg approach
extremely useful.  From such a correspondence
we can easily recognize the ADE type reductions of KP hierarchy
as the integrable structure of topological strings with the ADE type
potential \cite{DVV}. In a similar manner,
though they are a little bit involved, we can
identify certain reductions of Toda lattice hierarchy as the integrable
structure of the $c=1$ string theory \cite{DMP,GM,HOP,Tak,EK}
and the topological $CP^1$ string theory
together with its generalizations \cite{EY2,KO}.

In the language of local quantum field theory what we are looking at
is the deformation of topological action;
\eq
{\cal L}(t)~=~{\cal L}_0 + \sum_\alpha t_\alpha \int_\Sigma \phi_\alpha^{(2)}
+ \sum_{\alpha,n} t_{\alpha, n} \int_\Sigma \sigma_n( \phi_\alpha)^{(2)}~,
\en
where $\phi_\alpha^{(2)}$ and $\sigma_n( \phi_\alpha)^{(2)}$ are
two form observables related to the original zero form ones
$\phi_\alpha^{(0)}\equiv \phi_\alpha$ and $\sigma_n( \phi_\alpha)^{(0)}
\equiv \sigma_n( \phi_\alpha)$ by the descent equation;
\eqn
d \phi_\alpha^{(0)} &=& \Big [ Q_B, \phi_\alpha^{(1)} \Big]~, \CR
d \phi_\alpha^{(1)} &=& \Big [ Q_B, \phi_\alpha^{(2)} \Big]~.
\enn
In fact the topological conformal algebra enables us to solve
the descent equation in the following form;
\eq
\Phi^{(2)}~=~G_{-1} \overline{G_{-1}} \Phi^{(0)}~d^2 z~,
\en
for any zero form cohomology class $\Phi^{(0)}$.  In the Landau-Ginzburg
method the deformation of topological string may be described by the
perturbed potential. For example, we introduce
\eq
W(X,t)~=~\frac{1}{k+2} X^{k+2} + \sum u_i (t) X^i~,
\en
in the case of $A_k$ type potential. The perturbed potential is assumed to
satisfy
the condition \cite{DVV};
\eq
\Big\langle \phi_{I_1} \phi_{I_2} \cdots \phi_{I_n} \Big\rangle_{{\cal L}(t)}
{}~=~\Big\langle \phi_{I_1}(X,t) \phi_{I_2}(X,t) \cdots
\phi_{I_n}(X,t) \Big\rangle_{W(X,t)}~. \label{MIR}
\en
The left hand side is topological correlation function in the deformed
action ${\cal L}(t)$. In the right hand side one can compute the correlation
function by taking a summation over all the critical points of the potential
$W(X,t)$ \cite{Vaf};
\eq
\Big\langle \phi_{I_1}(X,t) \cdots
\phi_{I_n}(X,t) \Big\rangle_{W(X,t)}~=~\sum_{critical~points}~H^{g-1}(X,t)
\phi_{I_1}(X,t) \cdots \phi_{I_n}(X,t)~,
\en
where $g$ is the genus and $H$ is the Hessian of $W$.
Note that in the Landau-Ginzburg description the observables are defined
by
\eq
\phi_I (X,t) = \frac{\del}{\del t_I}~W(X,t)~,
\en
and, hence, they are $t$-dependent. This is consistent to the fact that
the topological BRS charge is also deformed according to
\eq
Q_B(t) = Q_B(0) - \sum t_I \oint \phi_I^{(1)}~.
\en
Thus in the Landau-Ginzburg description all the informations
of deformation are encoded in the perturbed potential $W(X,t)$.
The existence of $W(X,t)$ which satisfies the condition (\ref{MIR})
is a key to the integrable structure of topological string theory.
It seems that the secret of the relation (\ref{MIR}) is still not
well understood yet.

We compute the perturbed potential $W(X,t)$ by assuming a formal
power series expansion in the deformation parameters.
The coefficients of the expansion are obtained by estimating the (multi-)
contact terms. They are a result of contact term interactions between
topological observables. At the lowest order\footnote{The first derivatives are
the perturbed primaries by definition.}, the second
derivatives of $W(X,t)$ are given by the basic contact terms
$C(\phi_I, \phi_J)$;
\eq
\frac{\del^2 W}{\del t_I \del t_J} = \del_I \phi_J =
\del_J \phi_I  = C(\phi_I, \phi_J)~.
\en
In string theory we can implement the state-operator correspondence
by the path integral on the hemisphere with a fixed boundary condition
\cite{CV}.
Based on this correspondence,
we have the following interpretation of $\del_I\phi_J$;
\eq
\del_I \phi_J~=~\int_D \phi_I^{(2)} \vert \phi_J \rangle~,
\en
where $D$ is any small disk around the insertion point of $\phi_J$.
We can think of
the result of integration over $D$ as a result of contact interaction
of $\phi_I$ and $\phi_J$. The prescription of computing the
basic contact term $C(\phi_I, \phi_J)$ is as follows. We first
decompose the product $\phi_I\cdot\phi_J$ into two parts;
the first part is a linear combination of the primary fields and
the other part is (formally) BRS exact;
\eq
\phi_I\cdot\phi_J~=~C_{IJ}^\alpha \phi_\alpha + \Big[ Q_B,
\psi^- \lambda_{IJ} \Big]~.
\en
Then the contact term comes from the BRS exact part by the formula
\cite{CV,Los};
\eq
C(\phi_I, \phi_J)~=~\del_X \lambda_{IJ}~. \label{CON}
\en
Again the BRS exact term plays an important role in the game.
We can generalize this method of computation to the multi-contact
terms involving more than two operators.
The higher derivatives of $W(X,t)$ are expressed in terms
of these contact terms. For example the third derivatives are;
\eq
\frac{\del^3 W}{\del t_I \del t_J \del t_K} =
\Big( C(C(\phi_I,\phi_J), \phi_K) + \hbox {cyclic} \Big)
- C^{(3)}(\phi_I,\phi_J,\phi_K)~,
\en
where the last term is the higher contact term of three operators.
Thus the contact term interactions between physical operators
control the perturbed potential.

For the topological minimal models the perturbed superpotential is obtained
exactly in the small phase space where only the couplings $t_\alpha$ to the
primaries are turned on \cite{DVV}. The perturbed primary fields are given by
\eq
\phi_\alpha (X,t) = \frac{1}{\alpha+1}~\del_X \Big( L^{\alpha+1} \Big)_+~,
\quad (\alpha= 0,1, \cdots, k)~, \label{GEN}
\en
where $L$ is defined by the relation
\eq
W(X,t) = \frac{1}{k+2} L(X,t)^{k+2}~, \quad
L(X,t) = X + \cdots ~,
\en
and $(\cdot)_+$ means taking the non-negative power part.
It is apparent that the Laurent polynomial
$L$ plays the role of the Lax operator in the theory of integrable
system.  More precise treatment of $L$ as a Lax operator is given
in the theory of dispersionless integrable hierarchy \cite{Dub,TT}.
The leading term of $\phi_\alpha (X,t)$ is
$X^\alpha$ and the lower
order terms are developed by deformations. After coupling to
gravity the polynomials of higher power serve as the gravitational
descendants. Extrapolating (\ref{GEN}), we get \cite{EKYY};
\eq
\sigma_n(\phi_\alpha)~=~N_{n,\alpha} \del_X \Bigl[
L^{(k+2)n + \alpha} \Bigr]_+~, \label{GDS}
\en
where $N_{n,\alpha}$ is some normalization constant.
With the Lax like operator $L$,
this identification is natural in view of the property
that the gravitational descendants generate the higher
flows of the integrable hierarchy. Now the gravitational
descendants have the following \lq\lq Hodge
decomposition\rq\rq;
\eq
\sigma_n(\phi_\alpha)~=~\sum_\beta C_{(n,\alpha)}^\beta \phi_\beta
+ \del_X W \int^X \sigma_{n-1}(\phi_\alpha)~, \label{HOD}
\en
with the first term regarded as harmonic form.
By the formula of the contact terms (\ref{CON})
we immediately see the fundamental
recursion relation;
\eq
C(\phi_0, \sigma_n(\phi_\alpha)) = \sigma_{n-1}(\phi_\alpha)~. \label{PCN}
\en
We note that the primary field $\phi_0 = 1$ is identified
as the puncture operator $P$ after coupling to gravity.
Upon this identification the contact term (\ref{PCN}) implies
the puncture equation \cite{DW};
\eq
\Big\langle P \sigma_{n_1}(\phi_{\alpha_1}) \cdots
\sigma_{n_k}(\phi_{\alpha_k})
\Big\rangle ~=~\sum_{\ell =1}^k \Big\langle \sigma_{n_1}(\phi_{\alpha_1})
\cdots \sigma_{n_\ell -1}(\phi_{\alpha_\ell}) \cdots
\sigma_{n_k}(\phi_{\alpha_k}) \Big\rangle~.
\en
In deriving the puncture equation the first term of the decomposition
(\ref{HOD}) does not contribute. But the harmonic part is important
to obtain the following topological recursion relation at genus zero;
\eqn
\Big\langle \sigma_n(\phi_\alpha) \phi_\beta \phi_\gamma \Big\rangle
&=&\sum_{\delta=0}^k C_{(n,\alpha)}^\delta \Big\langle \phi_\delta
\phi_\beta \phi_\gamma \Big\rangle~, \CR
&=& \sum_{\delta=0}^k \Big\langle \sigma_{n-1}(\phi_\alpha)
\phi^\delta\Big\rangle
\Big\langle \phi_\delta \phi_\beta \phi_\gamma \Big\rangle~. \label{RCS}
\enn
The topological recursion relation (\ref{RCS}) has a clear
geometrical meaning on the moduli space of the Riemann
sphere with punctures \cite{W4}.
Thus our formula (\ref{GDS}) of the gravitational descendants
gives two basic relations among topological correlation functions.
It is known that (at genus zero) these recursion relations are equivalent
to the Virasoro constraint on the partition functions and
enough to fix it uniquely \cite{DVV2,FKN}.
In this sense the gravitational descendant obtained
as the $U(1)$ equivariant cohomology class is
the key to the integrable structure.
We think that the relation of the equivariant cohomology and
the integrability deserves further studies.

\section*{Acknowledgements}
I would like to thank the organizers of the symposium
for providing stimulating circumstances and the participants
for enlightening discussions.
This work is supported in part by the Grant-in-Aid
for Scientific Research from the Ministry of Education,
Science and Culture (No. 07210262).

\newcommand{\npb}[1]{{\it Nucl.~Phys.}~{\bf B#1}}
\newcommand{\mpla}[1]{{\it Mod.~Phys.~Lett.}~{\bf A#1}}
\newcommand{\cmp}[1]{{\it Commun.~Math.~Phys.}~{\bf #1}}
\newcommand{\plb}[1]{{\it Phys.~Lett.}~{\bf B#1}}
\newcommand{\ptp}[1]{{\it Prog.~Theor.~Phys.}~{\bf #1}}
\newcommand{\ijmp}[1]{{\it Int.~J.~Mod.~Phys.}~{\bf A#1}}
\newcommand{\lmp}[1]{{\it Lett.~Math.~Phys.}~{\bf #1}}
\newcommand{\jmp}[1]{{\it J.~Math.~Phys.}~{\bf #1}}

\end{document}